\documentstyle[12pt,epsf]{article}
\newcommand{\gsim}{ \raisebox{-.5ex}{\mbox{$\,\stackrel{>}{\sim}\,$}} }
\newcommand{\lsim}{ \raisebox{-.5ex}{\mbox{$\,\stackrel{<}{\sim}$\,}} }
\input{epsf}
\begin{document}


 \begin{center}
{\large\bf Poincar\'e Recurrences in Microtron and}\\
{\large\bf the Global Critical Structure}\\


         B.V. Chirikov\footnote{Email: chirikov@inp.nsk.su}\\ [5mm]
{\it Budker Institute of Nuclear Physics\\
 630090 Novosibirsk, Russia}
\end{center}


\begin{abstract}
The mechanism of the exponential transient statistics of Poincar\'e
recurrences in the presence of chaos border with its critical structure
is studied using two simple models: separatrix map and the kicked rotator
('microtron'). For the exponential transient to exist the two conditions 
have been shown to be crucial: fast (ballistic) relaxation, and a 
small measure of the critical structure.
The latter was found to include a new peripheral part (halo) of a surprisingly
large size. First preliminary empirical evidence is presented for a new
regime of Poincar\'e recurrences including the transition from exponential
to exponential statistics.
\end{abstract}


\baselineskip=24pt

\section{Introduction: exponential vs. power--law PR}
As is well known any trajectory of a bounded in phase space motion of
Hamiltonian system recurs infinitely many times to some neighborhood
of its initial position, for both regular (with discrete spectrum) as well as
chaotic (with continuous spectrum) motion. These Poincar\'e recurrences
(PR) do not imply a quasiperiodic motion which is still a widespread delusion
(see, e.g., \cite{1}). The difference between regular and chaotic motions
lies in the statistics of recurrences which is usually described by the
integral distribution $P(\tau )$ that is by the probability for a recurrence
time to be larger than $\tau$. In a regular motion 
such a survival probability $P(\tau )$
has a strict upper bound in $\tau$ while for a chaotic motion $\tau$ can be arbitrarily
long. In both cases PR characterize some fluctuations including arbitrarily
large ones in chaotic motion. The PR statistics proved to be a very powerful
and reliable method in the studies of chaotic dynamics due to its statistical
stability.

To my knowledge, such a method was first used (implicitly) in Ref.\cite{2}
for the study of a narrow chaotic layer along the separatrix of a nonlinear
resonance. The result ($\tau\geq 1$) 
$$
   P(\tau )\,\approx\,\frac{1}{\sqrt{\tau}} \eqno (1)
$$
was a surprise as it contradicted the bounded motion in chaotic layer.
Indeed, the total sojourn time $\tau\cdot P(\tau )$ of a trajectory, which 
is prportional to the
measure of the chaotic component of the motion,
diverges as $\tau\to\infty$.
Later \cite{3}, this apparent contradiction has been resolved simply
by increasing $\tau$ which showed that the exponent of the power--law decay
also increased from the initial $\nu =1/2$ to $\nu\approx 3/2$. 

It is instructive
to mention that the origin of a short--time computation in Ref.\cite{2}
was in apparently reasonable decision to avoid any rounding--off errors by
enormous increase of the computation accuracy. As a result, the computation
speed, and the available motion time, dropped by several orders of magnitude.
Generally, for exponentially unstable (chaotic) motion such an approach 
is prohibited whatever the computer power.
Fortunately, it is also unnecessary for calculating statistical characteristics
of the motion like $P(\tau )$ since most of the latter are robust.
True, the corresponding Anosov theorem \cite{4} was (and can be so far)
proved for the very simple Anosov systems only. Moreover, such a theorem is even
wrong for discontinuous (discrete) perturbations like rounding--off 
ones (see, e.g., Refs.\cite{5}).
Nevertheless, all the numerical experience confirms a sort of robustness
of the statistical behavior of chaotic systems, at least with some minimal
precautions (see, e.g., Refs.\cite{6} for discussion). Notice that without 
such an 'empirical' robustness
the numerical experiments with always {\it approximate} models would lose
any physical meaning!

A power--law decay $P(\tau )\sim\tau^{-\nu}$, whatever the exponent $\nu$,
found in \cite{2,3} for a bounded motion, was at variance with the exponential
decay believed to be a generic case. In Ref.\cite{3} the former was 
interpreted as a characteristic of a qualitatively new structure of the
motion near the chaos border in phase space. Later, it was termed the
critical structure, which was described by a renormalization group \cite{16} (see also
review \cite{7} and references therein).

Since then, the exponential decay has been considered as a property of
ergodic chaotic motion without any chaos borders. 
However, in recent numerical experiments \cite{8} with an asteroid motion
a fairly long transient exponential decay was found. Moreover, it persists
in the separatrix map also used, just the same map which seemed to have been 
well studied in many previous works \cite{3} (see also \cite{7} and
references therein). 

The main purpose of this paper is to reconsider various regimes of PR,
and to formulate the conditions for their realization using two relatively
simple models: separatrix and standard maps. Only bounded motion will be
considered, with or without chaos borders. First, a classical problem of PR
in an ergodic system will be discussed in some details in Section 2. Then, 
in Section 3, the analysis of various PR regimes in the separatrix map 
will be presented aimed to resolution of the apparent contradiction
mentioned above. In Section 4, PR in the standard map in accelerator
(microtron) regime will be described. The latter model presents a unique
possibility for quantitative study of the global critical structure. 
Particularly, a new part of this structure has been found which size
was surprisingly large. Finally, in Section 5, the main results of the present 
study are summarized. In addition, the first preliminary empirical 
evidence is presented for a new
regime of Poincar\'e recurrences including the transition from exponential
to exponential statistics.

\newpage

\section{PR in ergodic system: standard map}
Consider, first, an elementary example of 1D homogeneous
diffusion in momentum $p$. It can be described by a Gaussian distribution
function
$$
   f_G(p,t)\,=\,\frac{\exp{\left(-\frac{p^2}{2tD}\right)}}{\sqrt{2\pi tD}} 
   \eqno (2)
$$
where $D=<(\Delta p)^2>/t$ is the diffusion rate.
Derivative $f_P(p,t)=df_G/dp$ with boundary condition $f_P(0,t)=0$
which obeys the same diffusion equation
$$
   \frac{\partial f}{\partial t}\,=\,\frac{D}{2}\cdot\frac{\partial^2f}
   {\partial p^2} \eqno (3)
$$
describes, then, PR to $p=0$. The distribution of recurrence times (1) is
simply related to an auxiliary function $f_G$ by
$$
   P(\tau )\,=-\,A\int_0^{\infty}f_P(p,\tau )\,dp\,=\,Af_G(0,\tau )\,=\,
   \frac{A}{\sqrt{2\pi\tau D}}\,\approx\,\sqrt{\frac{\tau_0}{\tau\,+\,\tau_0}} 
   \eqno (4)
$$
Here $A$ is normalizing factor, and parameter $\tau_0$ provides a necessary
truncation of the preceding diverging expression at small $\tau$. 
It characterizes the dynamical time scale of the diffusion (cf., e.g., 
free path in molecular diffusion). If the motion in $p$ is actually
bounded (see below), Eq.(4) describes initial free diffusion.

\subsection{A little of theory}
Now, consider in more details another simple model -- the kicked 
rotator -- described by the so--called standard map:
$$
   \begin{array}{ll}
   \overline{p}\,=\,p\,+\,K\,\sin{x} & ({\rm mod}\ L) \\
   \overline{x}\,=\,x\,+\,\overline{p}\,-\,p_0 & 
   \end{array}
   \eqno (5)
$$
on a torus ($0\leq x<2\pi\,,\ 0\leq p<L=2\pi n,\ n=1,2,...$). 

We seek a solution $f(p,t)$ of diffusion equation (3) with the boudary 
condition 
$$
   f(0\,,t)\,=\,0 \eqno (6)
$$
which provides a loss of probability because of PR to $p=0$ (and to $p=L$).
The orthogonal and normalized eigenfunctions of the diffusion equation 
for this problem have the form ($k\geq 1$ is integer)
$$
   g_k(p)\,=\,\sqrt{\frac{2}{L}}\,\sin{\left(\frac{\pi kp}{L}\right)} 
   \eqno (7)
$$
with the corresponding eigenvalues
$$
   \gamma_k\,=\,\left(\frac{\pi k}{L}\right)^2\cdot\frac{D}{2} \eqno (8)
$$
which describe the decay rate of the eigenmodes (7). In Eq.(8) the diffusion
rate is
$$
   D(K)\,=\,\frac{K^2}{2}\cdot C(K) \eqno (9)
$$
with the dynamical correlation function \cite{9}
$$
   C(K)\,\approx\,1\,-\,2J_2(K)\,+\,2J_2^2(K) \eqno (10)
$$
where $J_2(K)$ is the Bessel function.

The set of eigenfunctions (7) and eigenvalues (8) provides a general solution
of the diffusion equation with boundary condition (6) for an arbitrary
initial distribution $f_0(p)=f(p,0)$. Peculiarity of PR statistics 
$P(\tau )$ is just
in a very particular initial condition. Specifically, for a single trajectory
in numerical experiments the recurrence time $\tau$ is determined by the
two successive crossings of the {\it exit line} which is, in the model under
consideration, $p=0\ {\rm mod}\ L$. 
Hence, the initial distribution is
concentrated right here: $|p|\leq K$. The condition for a trajectory
with initial $p>0$ to cross the exit line reads: $p+K\cdot\sin{x}<0$. Whence,
the probability of crossing is proportional to $\arccos{(p/K)}$, and the
normalized initial distribution can be taken in the form:
$$
   f_0\left(\frac{p}{K}\right)\,=\,\arccos{\left(\frac{p}{K}\right)}\,, 
   \quad 0\,<\,p\,\leq K\,<\,L \eqno (11)
$$
An example of $f_0$ is shown in the insert to Fig.1. It is convenient
to chose $f_0(p)$ on the one side of the exit line which is possible
due to the symmetry of eigenfunctions (7).

The difficulty with such an initial condition is in its narrow width 
which is always comparable with
the dynamical scale (both are $\sim K$, a single kick). 
This violates the diffusion
approximation for the exact integro--differential kinetic equation.
A simple remedy is well known, for example, from the theory of neutron
diffusion where the dynamical scale is the transport free path $l_n$
(see, e.g., Refs.\cite{10} and \cite{1}, p.689). A simple correction improving 
the diffusion
approximation amounts to a relatively small shift of the boundary condition 
(6) from
$p_b=0$ to $p_b=-\alpha l$ where $l$ is the dynamical scale
in our problem, and $\alpha\sim 1$ is unknown numerical factor to be
determined below from the numerical experiments. This implies an increase
of the global scale: $L=L_0+2\alpha l$ while the initial distribution
remains unchanged as it is obtained directly from the dynamics (5).
Notice the corresponding change in eigenfunctions (7).

The general solution of the diffusion problem is given by
$$
   f(p\,,t)\,=\,\sum_{k=1}^{\infty}\,f_k\cdot g_k(p)\cdot{\rm e}^
   {-\gamma_k\,t} \eqno (12)
$$
where the expansion coefficients $f_k$ are determined by the initial
condition (11):
$$
   f_k\,=\,\int_0^K\,\frac{dp}{K}\sqrt{\frac{2}{L}}\,
   \sin{\left(\frac{\pi kK}{L}\left(\frac{p}{K}\,+\,\Lambda\right)\right)}\cdot
\arccos{\left(\frac{p}{K}\right)}\,=\,
$$
$$
   \sqrt{\frac{\pi}{2L}}\cdot\frac{\cos{(sk\Lambda )}\cdot (1\,-\,{\rm J_0}
   (sk))\,+\,\sin{(sk\Lambda )}\cdot{\rm H_0}(sk)}{sk}\,\approx\, 
$$
$$   
   \frac{\pi sk}{4\sqrt{2L}}\,\left(1\,+\,\frac{8\Lambda}{\pi}\right)\cdot
   \left(1\,+\,{\cal O}(s^2k^2)\right)    \eqno (13)
$$
Here ${\rm H_0}$ is the Struve function, 
$\Lambda =\alpha l/K\sim 1$ (see below), and
$s=\pi K/L\ll 1$ is a small diffusion parameter. The latter
approximate expression in (13) holds true for $sk\lsim 1$.

Now, the PR statistics is described by
$$
   P(\tau )\,=\,\int_0^L f(p,\tau )\,dp\,=
   \sum_{k=1}^{\infty}\,f_k\,{\rm e}^{-\gamma_k\tau}\sqrt{\frac{2}{L}}
   \int_0^L\sin{\left(\frac{\pi kp}{L}\right)}\,dp\,=
$$
$$   
   =\,\frac{2}{\pi}\sqrt{2L}
   \sum_{m=1}^{\infty}\frac{f_k}{k}{\rm e}^{-\gamma_k\tau} \eqno (14)
$$
with $k=2m-1$ because only odd modes contribute to the integral.

Asymptotically, as $\tau\to\infty$, PR decay exponentially 
(Poisson statistics)
$$
   P(\tau )\,\to\,\frac{s}{2}\cdot\left(1\,+\,\frac{8\Lambda}{\pi}\right)\cdot
   {\rm e}^{-\gamma_1\tau}\,=\,F_1\cdot{\rm e}^{-\gamma_1\tau} \eqno (15)
$$
with the characteristic time
$$
   \tau_1\,=\,\frac{1}{\gamma_1}\,=\,\frac{2}{\pi^2}\cdot\frac{L^2}{D} 
   \eqno (16)   
$$
which is determined by the first (most slow) mode $m=k=1$, and which
is of the order of the global diffusion time.

The factor $F_1$ in Eq.(15) characterizes the share of asymptotic 
exponential decay which is small in the diffusive regime due to $s\ll 1$. 
The main, initial,
decay is a power--law one. Again, due to small $s$, the sum in Eq.(14) can be
approximately replaced by the integral over $m$ to obtain:
$$
   P(\tau )\,\approx\,\sqrt{\frac{\tau_0}{\tau}}\, 
             \approx\,\sqrt{\frac{\tau_0}{\tau\,+\,\tau_0}} \eqno (17)
$$
where 
$$
   \tau_0\,=\,\frac{\pi}{16\,C(K)}\cdot
   \left(1\,+\,\frac{8\Lambda}{\pi}\right)^2 \eqno (18)
$$
function $C(K)$ is given by Eq.(10), and the approximate expression
for $f_k$ in Eq.(13) is used.
The latter is not applicable for $\tau\to 0$, so that the final expression in
Eq.(17) is an approximate truncation of the preceding diverging relation
(cf. Eq.(4)).

The power--law/exponential crossover time $\tau_{cro}$ is obtained from
the comparison of
Eqs. (15) and (17), and is given approximately by the relation:
$$
   \tau_{cro}\,\approx\,\frac{L^2}{8\pi D}\,, 
   \qquad \gamma_1\tau_{cro}\,=\,\frac{\pi}{16} \eqno (19)
$$
Again, in the diffusive regime ($L^2\gg D$) the intermediate power--law decay
may be very long until the exponential asymptotics is reached.

\subsection{Numerical experiments}
An example of PR in ergodic case is shown in Fig.1. We use the standard map (5)
on a torus of sufficiently large circumference $L\gg K$ to provide a
diffusive relaxation ($s\ll 1$, for the opposite limit of ballistic
relaxation $s\gsim 1$ see Section 4 below). 
How strange it may seem,
the conditions for ergodicity even in such an apparently 'simple' model are 
still unknown ! However, numerical experiments (see, e.g., Ref.\cite{11} indicate 
that, at least,
for a particular value of the parameter $K=7$ the share of the regular domains, 
if any, is negligible ($\lsim 10^{-9}$) besides the two small islets 
(per map's 
period, see Section 4 below). Fortunately, their effect on PR is also 
negligible because they are related to the accelerator mode in which the momentum
$p$ quickly moves around the torus, so that a trajectory immediately crosses
the exit line $p=0\ {\rm mod}\ L$ (cf. Section 4).

\begin{figure}[]
\centerline{\epsfxsize=16cm \epsfbox{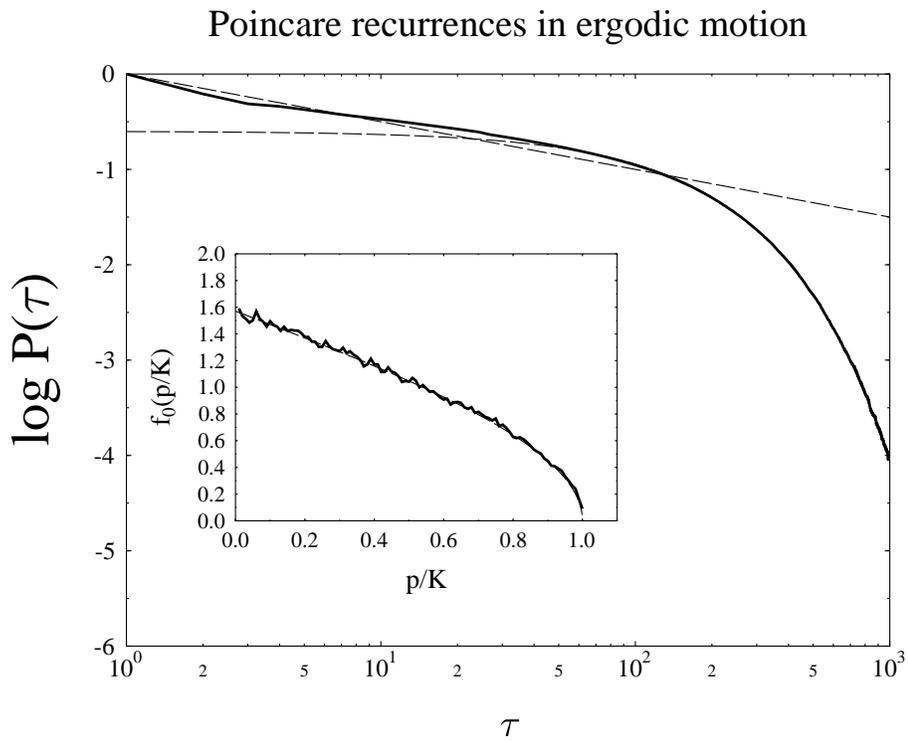}}
\caption{Poincar\'e recurrences in standard map (5) on a torus with
        exit line $p=0\ {\rm mod}\ L=50\pi$: $K=7$ ('quasiergodic' motion, 
        see text),
        $p_0=0$, a single trajectory of $t=10^7$ iterations.
        Solid line represents numerical data; dashed lines show two
        asymptotics: a power law (17) (straight line), and exponential (15).
        Insert: the initial distribution (11), just prior to crossing
        the exit line. Logarithms here and below are decimal.}
\end{figure}

In Fig.1 empirical data for a particular value of $L=50\,\pi$ are shown
which corresponds to $25$ periods of map (5) in $p$. All the data were obtained
from the run of a single trajectory over $10^7$ iterations.
Transition from a power law (straight dashed line) to an exponential (dashed
curve) is clearly seen. 

For a quantitative comparison with the theory above
(Section 2.1) we fix the dynamical parameter $l\equiv K\sqrt{C(K)}=
12.1$ where the value $C(7)\approx 3$ is used which has been obtained
from a special numerical experiment. It considerably differs from the
value $C(7)\approx 1.78$ according to approximate relation (10) just because
of accelerator islets mentioned above.
Since our model is a map, the minimal empirical recurrence time is 
$\tau_{em}=1$ instead of $\tau_{th}=0$ in a continuous theory
(for example, in numerical data $P(1)\equiv 1$).
The corresponding corrections are negligible except the initial dependence
for $\tau\sim 1$ (see below).

Numerical data in Fig.1 were fitted to Eq.(15) in the interval 
$\tau =500\ -\ 1000$ iterations, and the empirical values of the
characteristic time $\tau_1=125$, and of the factor $F_1=0.26$ were obtained. 
The corresponding values of the correction parameter are 
$\alpha_{\tau}=2.3$, and
$\alpha_F=0.71$. The difference in these two values of $\alpha$ characterizes
the accuracy of the correction which is rather poor because of a very narrow
initial distribution (see Eq.(11), and discussion around).
Without correction ($\alpha =0$) the theoretical values would be: 
$\tau_1=69$, and $F_1=0.07$ which both are substantially underestimated.

For a more systematic study the similar numerical data were computed for a
number of $L$ values specified by the integer $n=L_0/2\pi$. 
The results are shown in Fig.2.

\begin{figure}[]
\centerline{\epsfxsize=16cm \epsfbox{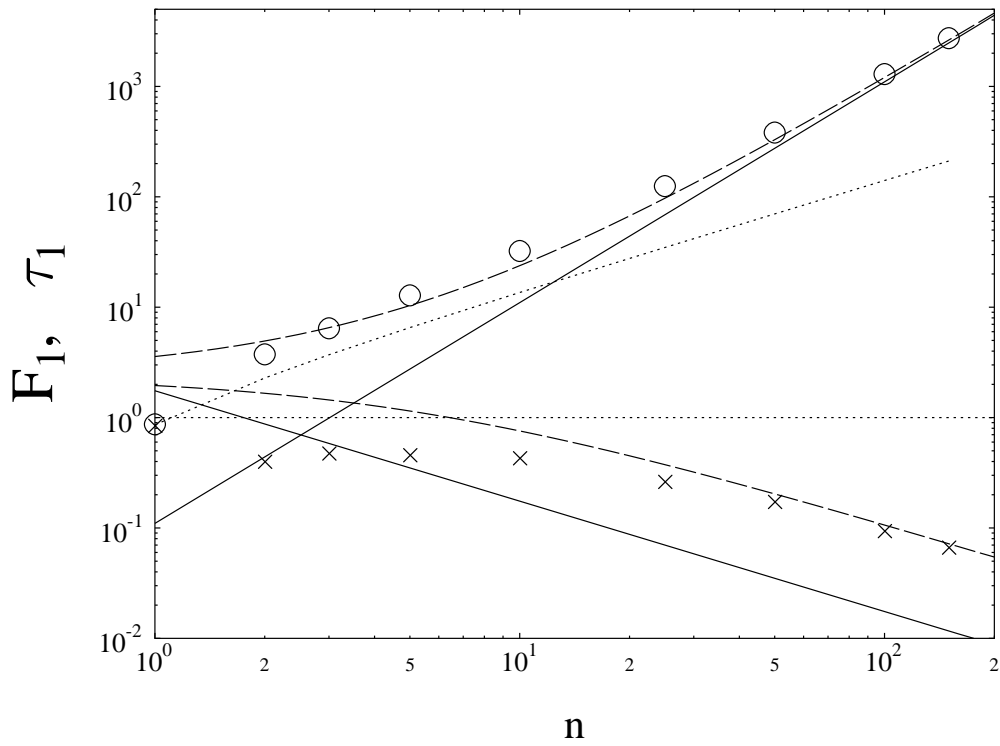}}
\caption{Asymptotic exponential in ergodic diffusive case:
         numerical values of characteristic decay time $\tau_1$ (16)
         (circles), and of
         the factor $F_1$ in Eq.(15) (crosses) vs. the torus size $n=L_0/2\pi$.
         Uncorrected dependence is shown by solid straight lines which are
         transformed into dashed lines by the correction with the same
         average $<\alpha >=1.22$. Two dotted lines represent the theoretical
         dependence (without any corrections) for the opposite, ballistic,
         limit, Eqs.(28,29)}
\end{figure}

Dependence $\tau_1(n)$ is well described by the uncorrected relation (16)
for large $n$ as expected. In intermediate region ($n\sim 3$) the agreement
is father improved by the correction which provides a smooth transition
to the ballistic limit (see Eqs.(28) and (29) in Section 4). In other words, 
the correction is not very important for the asymptotic decay rate 
because it is
determined by the first eigenfunction which is only slightly disturbed,
for large $n$, by the shift of the boundary. This is no longer the case for
the amplitude $F_1$ which strongly depends just on the distorted region
near the boundary $p=0$. As a result the correction is most important for
large $n$. The dependence $F_1(n)$ in the intermediate region remains unclear.
For $n\gsim 10,\ s\lsim 1/3$ both relations, Eqs.(15) and (16), are in a 
reasonable agreement with the numerical data for the same average value of the
correction parameter $<\alpha>=1.22$.

Coming back to Fig.1, we see that the initial power--law decay is well
described by a simple relation (17) with $\tau_0=1$ which is shown by the
dashed straight line, and which would correspond to $\alpha_0=0.66\approx
\alpha_F$.

\section{PR with a chaos border: separatrix map}
Now we consider an opposite limit of essentially nonergodic system
with a large chaos border and the critical structure.
As an example we take the separatrix map which was studied in many papers
(see, e.g., Refs.\cite{2,3,7}), and for which a new regime of PR 
has been recently
observed \cite{8}. The latter was the main motivation for the present studies.
We take the separatrix map in the form \cite{8}:
$$
   \begin{array}{ll}
   \overline{p}\,=\,p\,+\,\sin{x} & \\
   \overline{x}\,=\,x\,-\,\lambda\cdot\ln{(|\overline{p}|)}\,-\,p_0 & 
   \end{array}
   \eqno (20)   
$$
Here the motion is always strictly confined to the so--called chaotic
layer: $|p|\leq p_b(x)$. Previously, the most studied case corresponded
to big parameter $\lambda\gg 1$. In this limit $p_b\approx \lambda$, so that
the width of the layer ($2\lambda$) is much larger than the dynamical scale
of the diffusion (a single 'kick') which, for map (20), is unity (cf. Eq.(5)).
Besides the critical structure along the two borders, the average diffusion rate within the layer is nearly constant
(see Eq.(10)):
$$
   <D>\,\approx\,\frac{<C(\lambda /|p|)>}{2}\,\approx\,\frac{1}{2}
   \eqno (21)
$$
Hence, the initial decay of PR is a simple power law (1) which was observed, 
indeed, from the beginning \cite{2,3} (Section 1). The crossover time to a diferent 
law is given by a simple diffusion estimate:
$$
   \tau_{cro}\,\sim\,\frac{p_b^2}{D}\,\sim\,\lambda^2 \eqno (22)
$$
Unlike the ergodic case, the asymptotics of PR in the presence of chaos
border is also a power law but with a different exponent $\nu\approx 1.5$. 
This is explained by a very specific critical structure near the border
where the diffusion rate rapidly drops. As a result no trajectory can ever 
reach the exact border, even though it is approaching, from time
to time, the border arbitrarily close (see Refs. \cite{3,7,12} for details).

An example of this well known behavior is shown in Fig.3 (upper solid curve).
A transition between the two different power laws (dashed straight lines) 
at $\tau\sim 100$ is clearly seen in agreement with estimate (22).
There is no sign of any exponential decay. Now, how does it appear in a
similar model \cite{8}?

\begin{figure}[]
\centerline{\epsfxsize=16cm \epsfbox{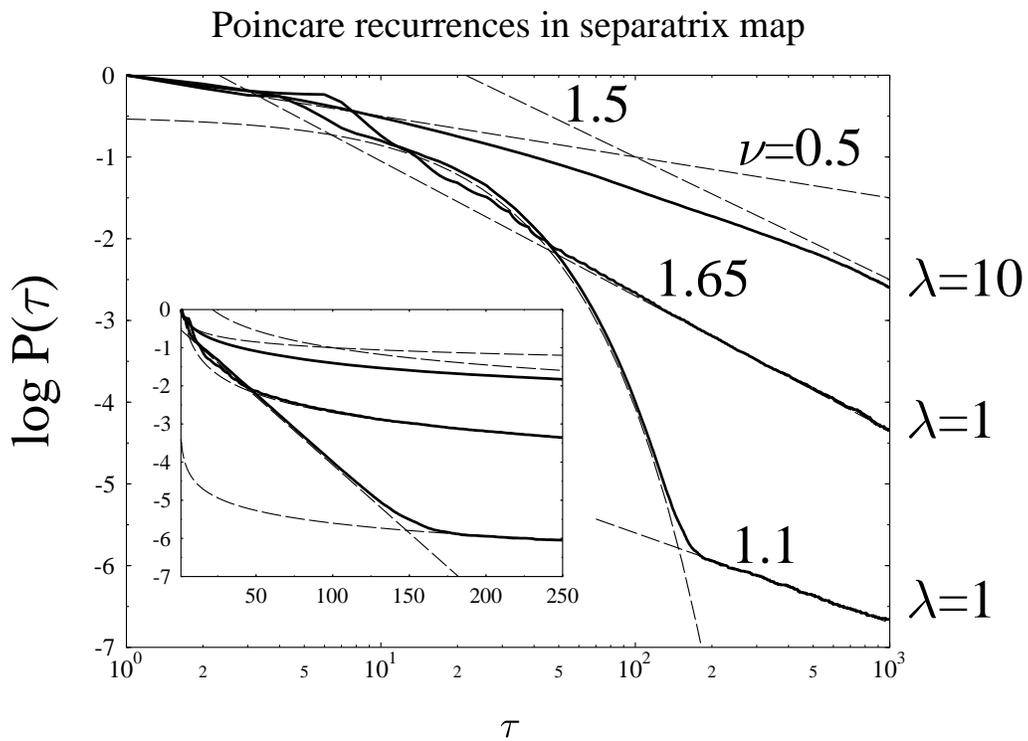}}
\caption{Poincar\'e recurrences in separatrix map (20): solid lines
         represent numerical data; straight dashed lines show the power
         law with the exponents $\nu$ indicated nearby; the values of parameter
         $\lambda =10\ {\rm and}\ 1$ are shown at the right; $p_0=0$.
         Two cases with the same $\lambda =1$ differ by the exit line
         (see text).
         Insert: the same in semi--log scale.
}
\end{figure}

The first observation is that in application to celestial mechanics
(dynamics of asteroids) the parameter $\lambda$ of map (20) is typically
rather small: $\lambda\sim 1$ \cite{8}. This drastically changes the structure
of the layer. First of all, the layer width is reduced down to the size of
a single kick. An example is shown in Fig.4. Hence, the diffusion approximation
becomes inapplicable. Instead, the so--called ballistic relaxation
comes into play which is much quicker. In other words, a slow diffusive
motion from the exit line to a critical structure is replaced now by rapid
jumps of a trajectory over the whole layer with some probability to get into
the critical structure. Since those jumps are very irregular in a chaotic
layer the PR 
are expected to decay exponentially. This is the case indeed as an example
in Fig.3 demonstrates (lower solid curve, $\lambda =1$ ). 
The exponential decay can be intermediate only as the trajectory is evetually
captured into the critical structure, and the decay turns to a power law.
Generally, the initial part of the power law is an approximate relation 
in that its exponent is
not universal, and is even varying with $\tau$. In the latter example
$\nu\approx 1.1$ which is rather different from $\nu\approx 1.5$ for the
upper curve in Fig.3.

Another interesting and important question is how long is the intermediate
exponential? For the lower curve in Fig.3 it is rather long: $\tau_{cro}
\approx 150$ which corresponds to the PR crossover as low
as $P_{cro}\approx 10^{-6}$ ! However, under different conditions with the
same $\lambda =1$ the exponential is much shorter: $\tau_{cro}
\approx 50$, and $P_{cro}\approx 10^{-2}$. The difference is in the exit line
as shown in Fig.4. 

\begin{figure}[]
\centerline{\epsfxsize=16cm \epsfbox{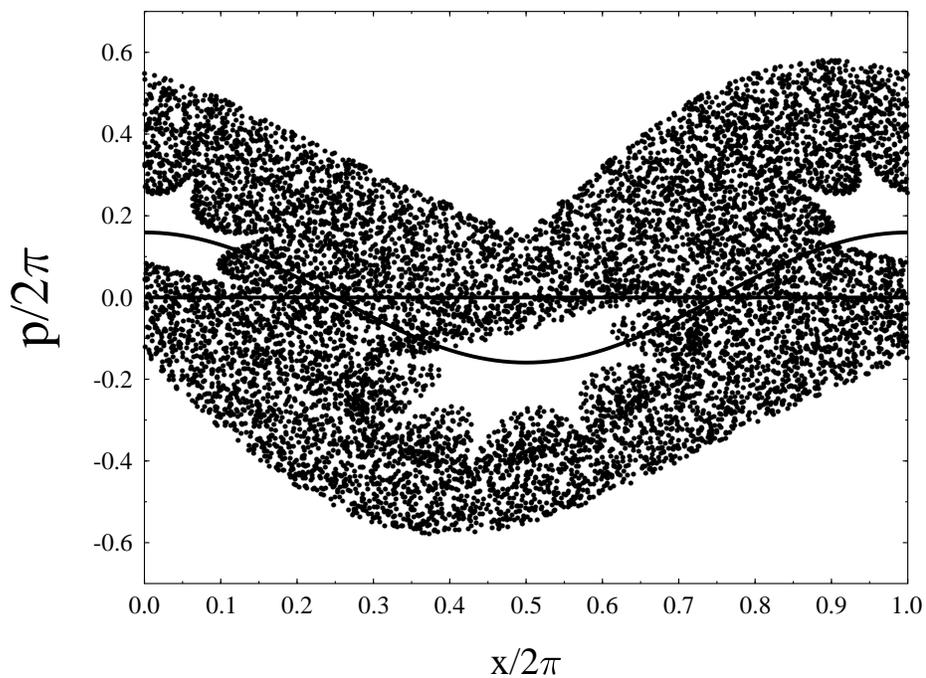}}
\caption{The phase space portrait of separatrix map (20) for $\lambda =1,\ 
p_0=0$. All points belong to a single trajectory of $10^4$ iterations.
The straight solid line is the usual exit line $p=0$  while another
one is the special exit line (23) which cuts through the two stability domains 
and thus blocks
the contribution of their big critical structure to PR.
}
\end{figure}

In the latter case the exit line is usual: $p=0$.
The critical structure is determined by the two big islands comparable in size
with that of the whole layer. This entails a rapid capture of a trajectory
into the critical structure, and a fast transition to a final power law
(with the local exponent $\nu\approx 1.65$). The lower curve in Fig.3 
corresponds to the same $\lambda =1$ but to a different exit line:
$$
   p_{ex}\,=\,\cos{(x)} \eqno (23)
$$
It is chosen in such a way to cut through both stability islands and, thus, 
to suppress any sticking to their critical structure. 
Then, the final power law is determined by the critical structure at the
layer borders which is apparently very narrow and cannot be discerned 'by eye'
in Fig.4. Nevertheless, it does exist as the asymptotic power law of PR 
in Fig.3 proves. Moreover, the latter even allows us to estimate 
the size of the
critical structure: its relative area (with respect to that of the layer)
is $A_{cr}\sim 4\times 10^{-4}$, or the width $(\Delta p)_{cr}\sim 10^{-3}$
(see Section 4).
This exponential transient is well fitted by the relation similar to (15)
(up to $\tau\approx\tau_{cro}$) with $\tau_1\approx 12$, and $F_1\approx 0.32$.
Both values are in a surprisingly well agreement with the uncorrected theory
($\alpha =0,\ L\approx 4.5$, see Fig.4) which gives $\tau_1\approx 8.2$, 
and $F_1\approx 0.35$. Apparently, this is because the diffusion parameter
$s\approx 0.7\sim 1$ is still not large enough.

Now, we can summarize the conditions for the transient exponential in PR
for a nonergodic motion: (i) fast, ballistic, relaxation, and (ii) a small
measure of the regular domains.
Besides, it turns out that the exponential PR allow for, at least, some
estimates of that measure. A more quantitative study of this interesting relation 
is convenient to continue with the standard map again. This is because
the latter has an infinite series of the special values of parameter
$K=K_n\approx 2\pi n$ for which there are well studied islands of regular 
motion with
a simple scaling and of rapidly decreasing area.

\section{PR in microtron: the standard map again}
The main advantage of this microtron model is in that it is very simple,
especially for numerical experiments, and well studied already.
Here we are interested primarily in the domains of regular motion which
exist for an infinite series of the special values of parameter
$K=K_n\approx 2\pi n$ where $n>0$ is any integer. 
Within these domains (islands) $|p|$ grows indefinitly
proportional to time which is the so--called microtron acceleration.
It was well studied since the celebrated paper due to Veksler in 1944
(see, e.g., \cite{13} and references therein). However, in the present paper, 
as well
as in Ref.\cite{13}, the main object for study is not the regular acceleration
itself but rather the chaotic motion outside the microtron islands which
is generally affected by the critical structure at the island borders.
A picture of this scale--invariant border is shown in Fig.5a
in dimensionless variables
$$
   x_s\,=\,(x\,-\,x_0)\cdot K\,, \quad 
   p_s\,=\,(p\,-\,p_0)\cdot K \eqno (24)
$$
where $p_0$ is a parameter of map (5), and
$$
   K\cdot\sin{x_0}\,=\,2\pi n\,, \quad K^2\,=\,\sigma^2\,+\,(2\pi n)^2\,, \quad
   \sigma\,=\,K\cdot\cos{x_0}\,, \quad -4\,<\,\sigma\,<\,0 \eqno (25)
$$
The latter inequalities determine the stability region around a fixed
point \newline $\pm x_0,\ p_0\ {\rm mod}\  2\pi$. In Fig.5a 
and below $\sigma =-2$ 
(the center of stability). 
For each integer $n$
there are two islets per phase space bin $2\pi\,\times\,2\pi$ one of which
is presented in Fig.5a. The picture shows a single trajectory of $5000$
iterations. During this time interval the trajectory is sticking to the
critical structure very close to the exact chaos border which results,
under particular conditions (see below), in an asymptotic power--law
decay of PR (cf. Fig.3 above). The island relative area (with respect to
that of the phase--space bin) is given also by a dimensionless relation \cite{13}:
$$
   A_n\cdot K_n^2\,=\,A_0(\sigma )\,\approx\,0.17  \eqno (26)
$$
where the latter value corresponds to $\sigma =-2$.
This area rapidly decreases as island's number $n$ grows.
Yet, for any $n\to\infty$ it determines the asymptotic PR decay, as we shall
see below.

\begin{figure}[]
\centerline{\epsfxsize=16cm \epsfbox{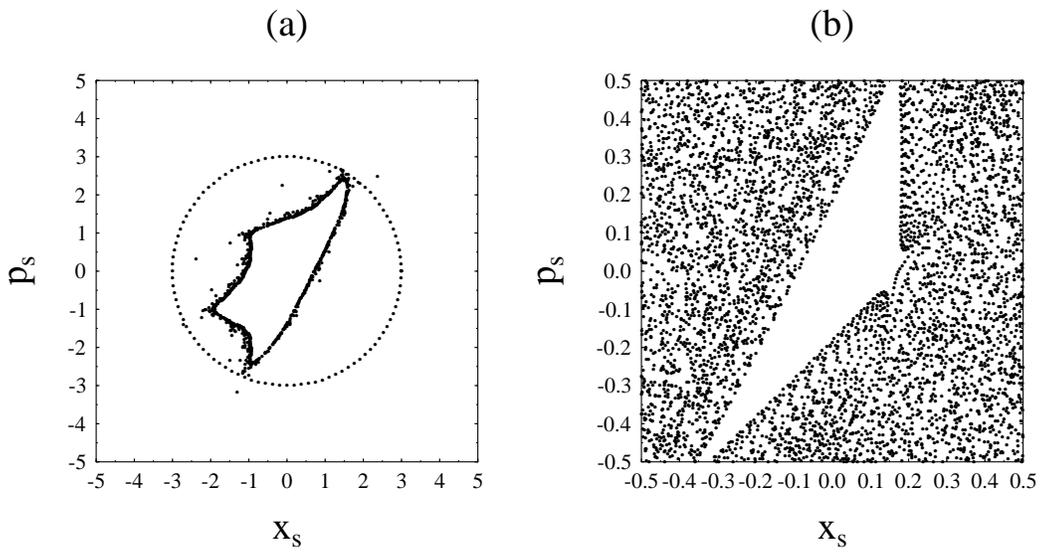}}
\caption{Universal border of microtron islets in scaled variables (24):
         (a) -- in the center of stability interval, $\sigma =-2$,
         single trajectory of $5000$ iterations stuck to the chaos border;
         (b) -- near the edge of stability, 
         $\sigma\approx -3.1$, revealed by an outside trajectory of
         $10^7$ iterations, $K=7$.
} 
\end{figure}

In Fig.5b another, much smaller, microtron island is shown for comparison.
In this case an outside, and much longer, trajectory was used which cannot 
ever cross the chaos border and enter the island. Its area is given by the
same estimate (26) with $A_0(-3.1)\approx 0.0038$.

The main difficulty with the microtron model for our purposes here is
the rapid growth in $|p|$ within and around the chaos border. This destroys
any long sticking of a trajectory whatever the exit line for PR (cf. Section 3,
Fig.4). To overcome this difficulty we used the following method. First, we
have chosen the exit line in such a way not to cut any island. It was done
simply by fixing parameter $p_0=\pi (\neq 0\ {\rm mod}\ 2\pi )$ in map (5)
without any change in the configuration of map's torus.
Second, we compensated acceleration by adding the term $2\pi n$ to the first 
Eq.(5). This helps, of course, for one island of each pair only.

Now, we need to provide the ballistic regime of relaxation that is a
sufficiently large parameter $s=\pi K/L$ (Section 2.1). It is convenient
to take $L=2\pi n$, so that the parameter
$$
   s\,=\,\frac{\pi K_n}{2\pi n}\,\approx\,\pi \eqno (27)
$$
is nearly independent of $n$ except a few small values of the latter.

Neglecting any dynamical correlations of the motion (particularly, those
caused by the presence of small microtron islets including the compensation
of acceleration) it is straightforward to
calculate the probability $w$, per map's iteration, for a trajectory to stay within the torus without
crossing the exit line. As is easily verified, it is given by the relation:
$$
   w\,=\,\int\frac{dx\,dp}{2\pi L/4}\,=\,\frac{2}{\pi L}\int_0^{x_m}\,dx\,
   (L\,-\,K\cdot\sin{x})\,=\,\frac{2}{\pi}\left(x_m\,-\,\frac{s}{\pi}
   (1\,-\,\cos{x_m})\right)\,\to\,
$$
$$
   1\,-\,\frac{2}{\pi}\,\approx\,0.363 \eqno (28)
$$
where
$$
   x_m\,=\,\left\{
   \begin{array}{cll}
   \arcsin{\left(\frac{\pi}{s}\right)}&, & s\geq \pi \\
   \pi /2 &, & s\leq \pi 
   \end{array} \right.
   \eqno (29)
$$
These general relations were used in Section 2.2 (Fig.2) to draw the
ballistic approximation.

The latter expression in Eq.(28) corresponds to the value $s=\pi$ 
used in numerical 
experiments. Without additional shift $\Delta p =2\pi n$ discussed above
the average time of the exponential decay would be
$$
   <\tau >\,=\,-\frac{1}{\ln{w}}\,\to\,0.988 \eqno (30)
$$
For $s=\pi$ the shift increases $w$ and $<\tau>$ up to
$$
   \tilde{w}\,=\,\frac{w}{2}\,+\,\frac{1}{2}\,=\,1\,-\,\frac{1}{\pi}\,\approx\,0.682
$$
$$
   \tilde{<\tau >}\,=\,-\frac{1}{\ln{\tilde{w}}}\,\approx\,2.61 \eqno (31)
$$
Now we can turn to numerical experiments with this microtron model.

\subsection{PR in microtron: numerics}
The main results of numerical experiments are presented in Fig.6, and
in the Table below. 
In Fig.6 the points show numerical data computed from a single
trajectory (for each $n$) up to $3\times 10^{11}$ iterations (for the
largest $n=5000$). The straight solid line is the fitted intermediate
exponential with the decay time $<\tilde{\tau}>=2.41$ in a good
agreement with the expected theoretical value $2.61$ in Eq.(31). This justifies
neglecting dynamical correlations assumed in the above 
theory in ballistic regime.

\begin{figure}[]
\centerline{\epsfxsize=16cm \epsfbox{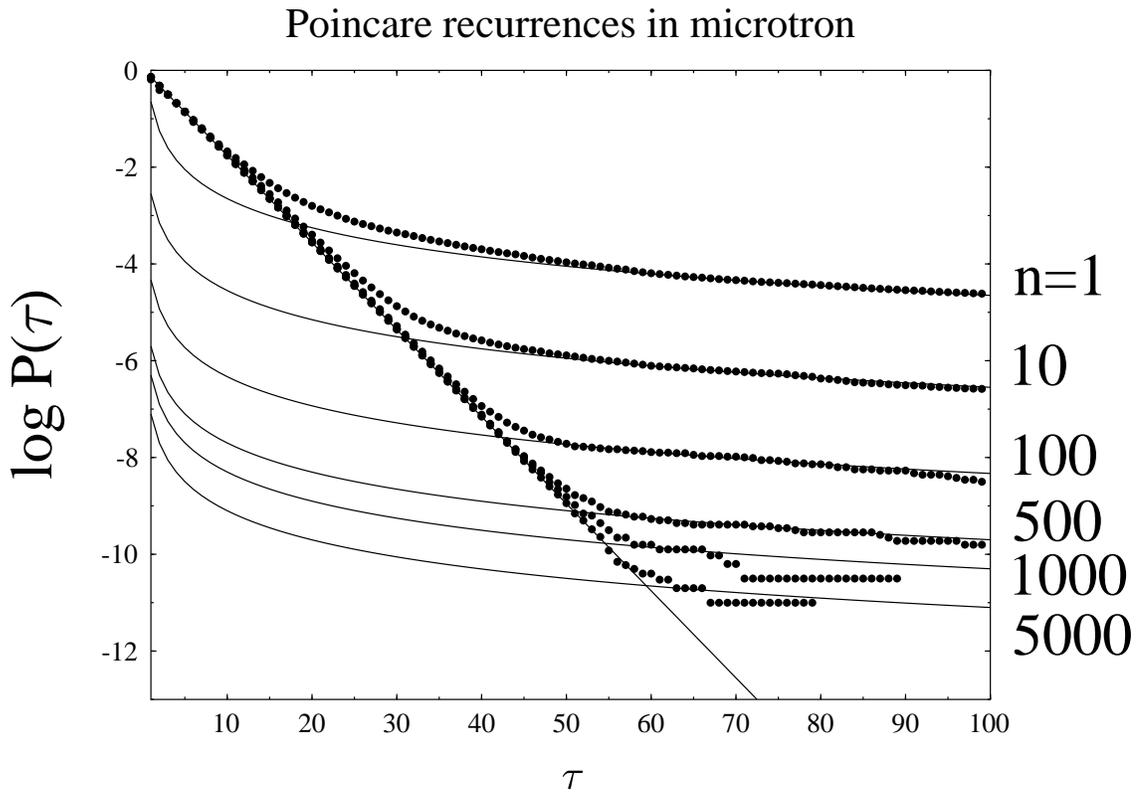}}
\caption{Poincare recurrences in microtron (points) for the values of the
microtron parameter $n$ given at the right. Straight line shows the 
intermediate
exponential with the fitted decay time $<\tilde{\tau}>=2.41$ (cf. Eq.(31)).
Solid curves give the fit of asymptotic power law (32) with nearly the same
exponent $\nu\approx 2$ extrapolated back to $\tau =1$ (see text).
} 
\end{figure}

The exponential/power--law crossover time systematically increases with
$n$ that is with the decrease of the microtron island area (see Table).
The power--law tails of PR were fitted by the expression
$$
   P_n(\tau )\,=\,\frac{A_{cr}(n)}{\tau^{\nu}} \eqno (32)
$$
Remarkably, all values of the exponent were found to be close: $\nu\approx 2$.
The relation of this expression to the size of the critical
structure is based on the following hypothesis:
dependence (32), fitted to the tail of PR, can be extrapolated
back to $\tau =1$. If true, it allows us to interpret
the parameter $A_{cr}(n)$ as the relative area of the whole (global) critical 
structure around the corresponding microtron island of area $A_n$.

One could expect that both areas are comparable: $A_{cr}(n)\sim A_n$.
Surprisingly, this is not the case (see Table, third column; the data in
fourth column will be discussed below). Their ratio
$R=A_{cr}/A_n\sim 100$ is not only very large but also slowly 
increasing with $n$ according to the following approximate
empirical relation 
$$
   R(n)\,\approx\,50\,n^{1/4} \eqno (33)
$$
The origin of this small correction to a simple scaling $R\approx$ const
remains unclear.

\vspace{1cm}
{\large
\begin{center}
\begin{tabular}{|c|l|c||c||}
\multicolumn{4}{l}{Table. Global critical structure around microtron islets}\\ [3mm] \hline
\multicolumn{1}{|c|}{$n$} &
\multicolumn{1}{|c|}{$A_n$} &
\multicolumn{1}{|c||}{$R$} &
\multicolumn{1}{|c||}{$R_{ex}$} \\ 
islet number & islet area & Fig.6 & Fig.7 \\ \hline
1 & $4.30\times 10^{-3}$ & 36 & 25.0 \\
5 & $1.71\times 10^{-4}$ & 124 & -- \\ 
10 & $4.31\times 10^{-5}$ & 65 & 13.9 \\
100 & $4.31\times 10^{-7}$ & 202 & 10.6 \\
500 & $1.72\times 10^{-8}$ & 291 & -- \\
1000 & $4.31\times 10^{-9}$ & 176 & 9.8 \\
5000 & $1.72\times 10^{-10}$ & 461 & -- \\ \hline
$10^4$ & $4.31\times 10^{-11}$ & -- & 8.6 \\
$10^5$ & $4.31\times 10^{-13}$ & -- & 8.3 \\
$10^6$ & $4.31\times 10^{-15}$ & -- & 8.1 \\
$10^7$ & $4.31\times 10^{-17}$ & -- & 7.7 \\ \hline
\end{tabular}
\end{center}
}

In any event, the size of the whole critical structure seems to be 
much larger than expected. This main outer part of the structure looks
ergodic, and forms a sort of halo around the usually narrow inner part
with a typical admixture of chaotic and regular components of motion.
The former reminds the ergodic critical structure around a parabolic 
fixed point,
that is the limiting case of an island of zero size, studied in Ref.\cite{14}.
In a sense, such a halo is some 'hidden' critical structure,
without internal chaos borders but with apparently strong
correlations in the motion which keep a trajectory within this relatively 
small domain.

Now, the principal question to be aswered reads: is the observed halo a real
physical structure or the result of a wrong interpretation of 
the empirical data
using the above extension hypothesis?

\subsection{Exit time statistics}
To clarify this question a new
series of numerical experiments was undertaken.
To this end, the exit times from the halo, instead of 
recurrences, were measured.  Such a method was recently successfully used in the
studies of the critical structure in Ref.\cite{15}.
In the problem under consideration here the measurement of exit
times was organized as follows.
A number (typically 100) of trajectories with the initial conditions
homogeneously distributed over the circle around a microtron island
(see Fig.5a) were run until they leave the interval 
($0<x<\pi$). The dependence of the 
average exit time $\tau_{ex}$ for a series of the circles with increasing
radius $\rho_s$ as a function of the area within a circle $A_s=\pi\rho_s^2$
(in scaled variables (24)) was thus computed. The minimal circle of radius 
$\rho_s=3$ touches
the island, and comprises the area $A_{min}=28.3$ while island's area in
these units is $A_{sn}=6.72$, the minimal ratio being
$R_{ex}=A_s/A_{sn}=4.21$.

The main results of this measurement are shown in Fig.7 for 8 different
values of parameter $n$ up to $n=10^7$ with the island area as small as
$A_n\approx 4\times 10^{-17}$ ! This is completely out of reach 
for the PR method
(cf. Ref.\cite{15}). The difference is in a rather short exit time from the
halo, we are interested in, as compared to the long recurrence time on the
tail where it is eventually separated from the exponential (Fig.6).

\begin{figure}[]
\centerline{\epsfxsize=16cm \epsfbox{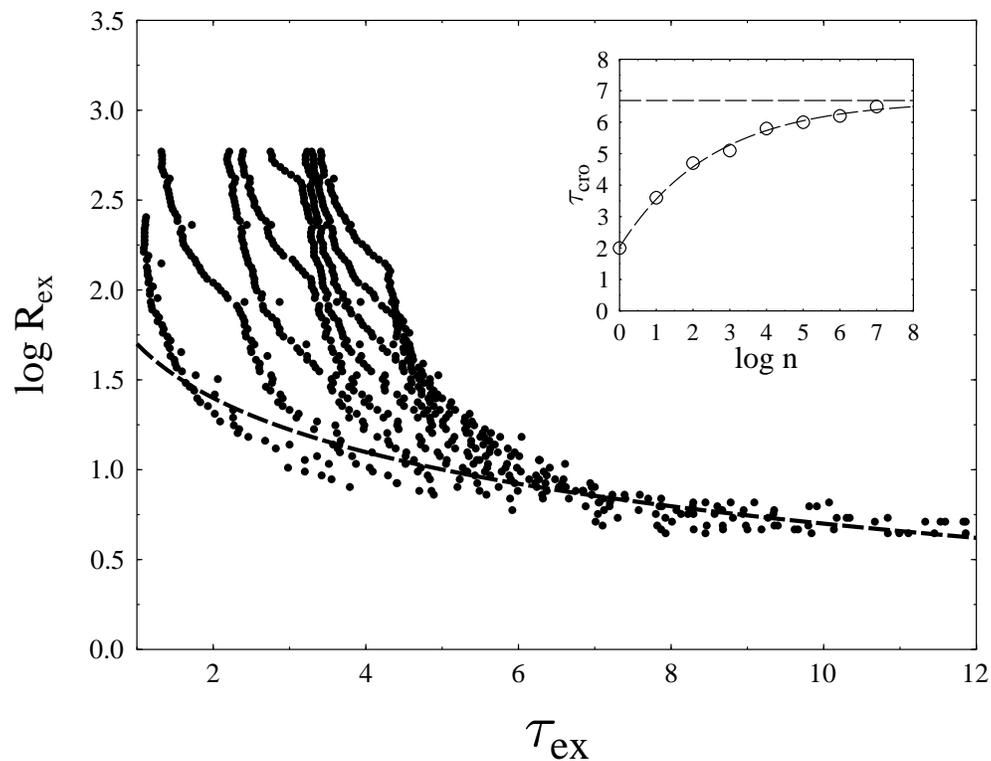}}
\caption{Scaling of exit time from halo: the ratio $R_{ex}$ of the critical 
         to island area vs. the average exit time out of the
         former (points) from a hundred trajectories (for each of 8 values
         of $n$) with initial
         conditions on a series of circles (see Fig.5a and the text).
         Dashed line shows the standard critical scaling (34) which breaks down
         at some values of crossover time $\tau_{cro}$ depending on $n$
         (insert).
} 
\end{figure}

The main result revealed in Fig.7 is a transition between the two different
scalings. One, for relatively large $\tau_{ex}$, is the standard critical
scaling shown by the dashed line which is the fitting of numerical data
to the relation
$$
   R_{ex}(\tau_{ex})\,\approx\,\frac{R_{ex}(1)}{\tau_{ex}} \eqno (34)
$$
with $R_{ex}(1)\approx 50$.
As expected, this part of the data does not depend on $n$. Moreover, scaling
(34) is in a good agreement with the PR tail in Fig.6.
The relation between the two is well known \cite{3,7,13,15}.
Generally, the power--law PR statistics is descibed by (cf. Eq.(32)):
$$
   P(\tau )\,\approx\,\frac{P(1)}{\tau^{\nu}}\,\approx\,
   A(\tau )\,\frac{<\tau >}{\tau}
   \eqno (35)
$$
where $<\tau >$ is the average PR time, and the latter expression is
obtained from the ergodicity within the chaotic component of the motion.
Using approximate relation $\tau\approx 2\,\tau_{ex}$ \cite{15}
we obtain
$$
   R_{ex}(1)\approx\,{A(2)\over A_n}\,\approx\,{R\over <\tau >\,2^{\nu -1}}
   \eqno (36)
$$
where $R=A_{cr}/A_n\approx A(1)/A_n$ (see above).
For integer map's time
$$
   <\tau >\,=\,\sum_{\tau =1}^{\infty}\tau\left(\frac{1}{\tau^{\nu}}\,-\,
   \frac{1}{(\tau +1)^{\nu}}\right)\,=\,\zeta (\nu ) \eqno (37)
$$ 
where $\zeta$ is the Riemann function. Whence,
for $\nu =2$ the relation (36) gives (see Eq.(34))
$$
   R\,\approx\,\frac{\pi^2}{3}\,R_{ex}(1)\,\approx\,160 \eqno (38)
$$ 
which is in a reasonable agreement with numerical data (third column in Table).
However, unlike the data in Fig.6 where the actual power--law scaling is not
seen under much larger exponential transient, the data in Fig.7 clearly
demonstrate that the critical scaling does not reach the limit
$\tau =1$ assumed above. Moreover, the crossover time $\tau_{cro}$ 
increases, and hence
the size of the global critical structure ($R_{ex}$) decreases, as $n$ grows
(Fig.7, insert). The increase of $\tau_{cro}$ must have an upper bound because
otherwise the critical structure near the chaos border would be also destroyed
in contradiction to the detailed studies of that in anomalous diffusion
\cite{13}. Indeed, the empirical dependence $\tau_{cro}$ in Fig.7 (insert)
can be fitted reasonably well by the expression
$$
   \tau_{cro}\,=\,6.69\,-\,\frac{4.66}{n^{0.172}} \eqno (39)
$$
The upper limit in $\tau_{cro}$ corresponds, according to Eq.(34),
to the lower limit in the ratio $R_{ex}>7.47$. Combining Eqs. (34) and (39)
we obtain approximately
$$
   R_{ex}(n)\,\approx\,\frac{R_{ex}(1)}{\tau_{cro}(n)} \eqno (40)
$$
The empirical values of this dependence are given in Table (fourth column).
They indicate much smaller, yet still a fairly large, size of the critical 
structure as compared
to the limiting estimate for PR (third column). The former seems to be
more reliable and realistic.
A different, new and unknown, scaling in Fig.7 for $\tau_{ex}<
\tau_{cro}$ requires further studies. What is of importance here
is the termination of the critical scaling at a finite $\tau_{ex}
=\tau_{cro}$. This determines the outer border of the critical structure.

\section{Discussion: a new puzzle}
The original motivation of these studies was the unusual exponential transient
observed in PR in the presence of chaos border \cite{8}.
However, in the course of investigating the mechanism and conditions 
of this phenomenon a more interesting observation has come out. It suggests
the existence of a new, unknown to my knowledge, part of the critical
structure surrounding, like a halo, the well--known inner part close to
the chaos border. In spite of some contradictory empirical evidence
the halo apparently occupies the most of the global critical structure. 
In any event, in the microtron model considered in this paper the area
of the halo is much larger than that of the regular island inside it,
even according to the minimal estimates (see Table and Fig.7).

As is well known, the scaling of the peripheral part of the critical structure
is generally nonuniversal, at least quantitatively, in the sense of 
the corresponding
power--law exponents, for example \cite{15}.
However, it might be nevertheless typical qualitatively as it appears
in our model.
In this respect, it would be interesting to look at different examples 
of the global 
crtical structure. One possibility is to use the same model with a fixed
parameter $L=2\pi n$ (Section 4) but for different values of the 
stability parameter $\sigma$ in
Eq.(25). First preliminary numerical experiments have been done for 9
values of $\sigma$ within the whole stability interval ($-4\leq\sigma\leq 0;\ 
2\pi\leq K\leq 7.45$) including the 'quasiergodic' case $K=7$ used in
Section 2 for other purposes. In all cases but the latter the PR behavior
was similar to that in the main series of numerical experiments (Fig.6),
at least qualitatively. However, just for $K=7$ a sudden surprise has emerged
which is presented in Fig.8. 

\begin{figure}[]
\centerline{\epsfxsize=16cm \epsfbox{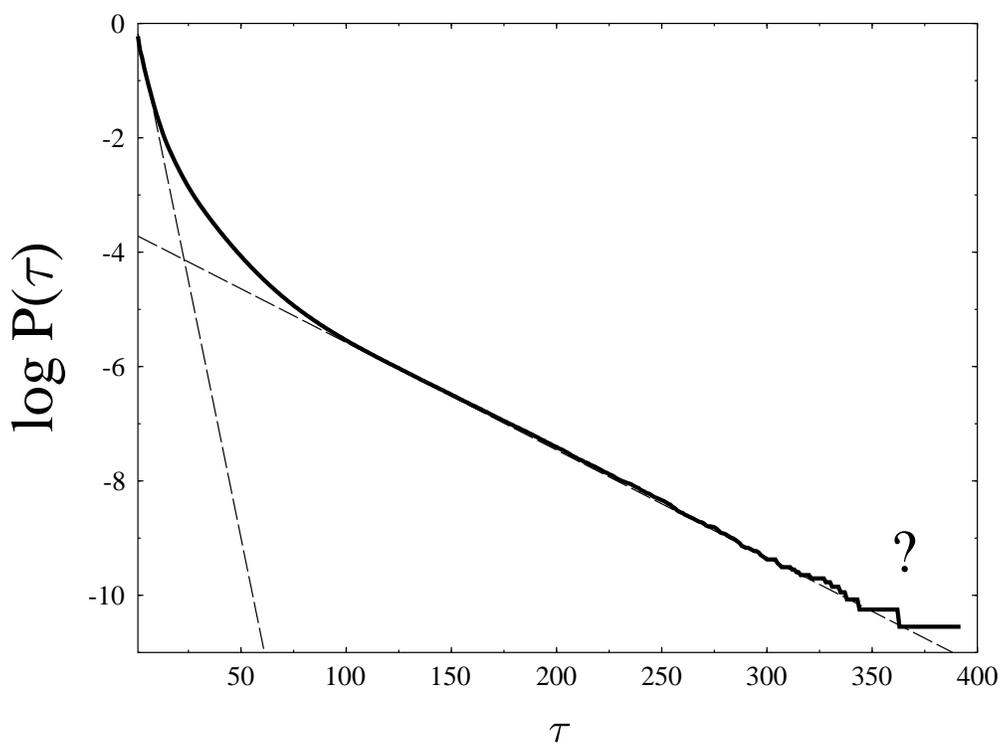}}
\caption{Poincar\'e recurrences for $L=2\pi$, and $K=7$. The solid line shows
         numerical data from a single trajectory of $10^{11}$ iterations. 
         Two dashed lines are fitted exponentials with the average decay time
         $<\tilde{\tau}>=2.41$, and $<\tilde{\tau}>=23.1$, respectively.
} 
\end{figure}

In spite of a very long run ($10^{11}$ iterations) no clear sign of the
expected power--law decay is seen. A small deviation from the final exponential
at the end of the dependence is a typical feature due to a poor statistics
(cf., e.g., Fig.6). The first exponential is close to the expected one with
the fitted decay time $<\tilde{\tau}>=2.41$ as compared to the theoretical
$<\tilde{\tau}>=2.38$ (see Section 4). For the second exponential the empirical
decay time $<\tilde{\tau}>=23.1$ is about 10 times longer. This means that a
trajectory is kept within (sticks to?) a certain domain but not in a way
it does so in the usual critical structure. Moreover, the relative area 
$A_d\sim 2\times 10^{-4}$ of this peculiar domain, estimated similarly to
$A_{cr}(n)$ in Eq.(32), is small and is comparable with that of the island 
inside (Fig.5b): $A_7\approx 7.8\times 10^{-5}$. This island does have
a chaos border, yet contrary to usual behavior, it does not produce any
appreciable power--law decay of PR. Another preliminary remark is that
a more careful inspection of Fig.5b seems to suggest a different, more
regular than usual, structure of the chaos border for $K=7$ (cf. Fig.5a).
Certainly, this 'anomaly' deserves further investigation.

\bigskip

{\bf Acknowledgements.} I am grateful to I.I. Shevchenko for many interesting discussions and important remarks.
This work was partially supported by the Russia
Foundation for Fundamental Research, grant 97--01--00865.


\end{document}